\documentclass[twocolumn,english,prb]{revtex4}
\usepackage[latin1]{inputenc}
\usepackage{amsmath}
\usepackage{amssymb}

\makeatletter

\usepackage{graphicx}
\usepackage{amssymb}

\usepackage{revsymb}

\renewcommand{\vec}[1]{\mathbf{#1}}

\renewcommand*{\sectionmark}[1]{} 
\renewcommand*{\subsectionmark}[1]{} 

\usepackage{babel}

\usepackage{babel}
\makeatother
\begin{document}
\newcommand{\ket}[1]{|#1\rangle}
\newcommand{\bra}[1]{\langle#1|}
\newcommand{\braopket}[3]{\left\langle #1\left|#2\right|#3\right\rangle }
\newcommand{\bsigma}{\boldsymbol{\sigma}}
\newcommand{\brho}{\boldsymbol{\rho}}
\newcommand{\bzeta}{\boldsymbol{\zeta}}
\newcommand{\bvarepsilon}{\boldsymbol{\varepsilon}}
\newcommand{\ShermanBmAvg}{\gamma}
\newcommand{\EFermi}{\varepsilon_{\mathrm{F}}}

\newcommand{\SOcoupling}{\lambda}

\newcommand{\unitVector}[1]{\hat{\vec{#1}}}

\newcommand{\jCharge}{\vec{J}_{\mathrm{c}}}
\newcommand{\jChargeComp}[1]{J_{\mathrm{c}}^{#1}}

\newcommand{\jCCC}[3]{j_{#1,\,#3}^{#2}}

\newcommand{\jSHc}{j}
\newcommand{\jSHSJ}{\jSH_{\,\mathrm{SJ}}}
\newcommand{\sigmaSHSS}{\sigma_{\mathrm{SS}}^{\mathrm{SH}}}
\newcommand{\jS}{\vec{j}_{\mathrm{s}}}

\newcommand{\vdrift}{\vec{v}_{\mathrm{dr}}}
\newcommand{\vdriftAbs}{\mathrm{v_{dr}}}
\newcommand{\Ldrift}{L_{\mathrm{dr}}}

\newcommand{\tPulse}{t_{\mathrm{p}}}

\newcommand{\jSz}{\vec{j}_{\mathrm{s}}^{z}}
\newcommand{\jSHz}{\vec{j}_{\mathrm{SH}}^{z}}
\newcommand{\jSzc}[1]{j_{#1}^{z}}
\newcommand{\divergence}{\mathrm{div}\,}
\newcommand{\spingen}{\Gamma_{z}}
\newcommand{\surfRelax}{k}

\newcommand{\taus}{\tau_{\mathrm{s}}}
\newcommand{\Ls}{L_{\mathrm{s}}}
\newcommand{\Ds}{D_{\mathrm{s}}}

\newcommand{\crossSecSpin}{d\negmedspace\!\stackrel{\leftrightarrow}{\sigma}\negmedspace\!}

\newcommand{\conductivityTensor}{\hat{\sigma}}
\newcommand{\SHtensor}{\hat{\sigma}^{\mathrm{SH}}}
\newcommand{\AHtensor}{\hat{\sigma}^{\mathrm{AH}}}
\newcommand{\SHdiagonal}{\hat{\sigma}_{\mathrm{d}}^{\mathrm{SH}}}

\title{Spin generation away from boundaries by nonlinear transport }

\author{Ilya G. Finkler, Hans-Andreas Engel, Emmanuel I. Rashba, Bertrand
I. Halperin}

\affiliation{Department of Physics, Harvard University, Cambridge, Massachusetts
02138 }

\begin{abstract}
In several situations of interest, spin polarization may be generated
far from the boundaries of a sample by nonlinear effects of an electric
current, even when such a generation is forbidden by symmetry in the
linear regime. We present an analytically solvable model where spin
accumulation results from a combination of current gradients, nonlinearity,
and cubic anisotropy. Further, we show that even with isotropic conductivity,
nonlinear effects in a low symmetry geometry can generate spin polarization
far away from boundaries. Finally, we find that drift from the boundaries
results in spin polarization patterns that dominate in recent experiments
on GaAs by Sih \emph{et al.} {[}Phys. Rev. Lett. \textbf{97}, 096605
(2006)]. 
\end{abstract}


\maketitle
Spin polarization can be generated and manipulated in semiconductors
by means of electric fields and spin-orbit coupling. A prominent example
is the spin Hall effect, \cite{dp71,hirsch,kato,murakami,sinova,wunder,ehr,tds,ehrReview}
where a homogeneous electric current passing through a sample induces
spin polarization $\vec{s}(\vec{r})$ near lateral edges, with opposite
polarization at opposite edges. For rectangular homogeneous samples,
this spin polarization falls off exponentially when moving away from
the boundary on the length scale of the spin diffusion length $\Ls$,
so for samples of size $L>\Ls$ no spin polarization due to the spin
Hall effect is expected far away from the edges (on scale $L$). However,
in recent experiments on low-symmetry samples, Sih \textit{et al.}
\cite{sih} observed polarized spins away from the edges of a GaAs
sample subjected to an electric current and concluded that there are
transport effects beyond the simplest spin Hall effect near edges.

In existing analytical theories of the spin Hall effect, spin polarization
in spin-orbit coupled media has been considered in the linear transport
regime for small electric field $\vec{E}$. Here we show that in this
regime, spin generation away from boundaries is forbidden or strongly
suppressed for sufficiently symmetric samples. However, in the context
of extrinsic effects, by considering nonlinearities in charge transport,
we find a new mechanism of generating electron spins by electric current.
This nonlinear regime is of practical importance, because experiments
are often performed in a range of electric fields with nonlinear current-voltage
characteristics.~\cite{kato,davidprivate} We then present an analytically
solvable model in which the bulk spin generation in a radially symmetric
sample geometry is due the nonlinearity and the anisotropy of conductivity
tensor. Finally, we numerically solve the charge transport and spin
drift-diffusion equations for the sample geometries used in Ref.~\onlinecite{sih}.
The patterns of spin accumulation we find strongly resemble experimental
findings. We establish the existence of two contributions to the spin
polarization: the first contribution is generated at the boundaries
and then drifts large distances (while diffusing further away from
the boundaries), and the second one is generated away from boundaries.
While we find that in the recent experiments the first of these dominated,
we propose setups that should allow unambiguous observation of the
spins generated in the bulk.

We consider a diffusive system with a weak extrinsic spin-orbit coupling
and typical system size $L$.  The spin current contains drift and
diffusion contributions. Additionally, electrical current, via the
extrinsic spin-orbit coupling, induces a spin Hall current $\jSHz$,
which acts as a source for the spin polarization $s_{z}(\vec{r})$
 with a generation rate $\spingen=-\divergence\jSHz$. For length
scales much larger than the mean free path $\ell$,  spin-orbit processes
(e.g., Dyakonov-Perel or Elliot-Yafet mechanisms) lead to a finite
spin lifetime $\taus$ . Together with a spin diffusion coefficient
$\Ds$, the spin lifetime defines a spin diffusion length $\Ls=\sqrt{\Ds\taus}$.
 To analyze the spin polarization away from boundaries, we consider
the regime $\ell\ll\Ls<L$. On length scales large compared to $\ell$
and in the absence of an external magnetic field, the spin density
$s_{z}(\vec{r})$ obeys the drift-diffusion equation \begin{equation}
\dot{s}_{z}=\divergence\left(\Ds\nabla s_{z}\right)-\divergence\left(\vdrift s_{z}\right)+\spingen-\frac{s_{z}}{\tau_{s}},\label{eq:kinetic}\end{equation}
 where drift velocity $\vdrift$ is proportional to the local electric
current density. This introduces yet another length scale, the drift
length $\Ldrift=\vdriftAbs\taus$.

In the linear transport regime, one can evaluate $\spingen$ by writing
the spin current induced by the electric field as $\jSHz=\SHtensor\,\vec{E}$,
with the spin Hall conductivity tensor $\SHtensor$ that does not
depend on $\vec{E}$. Assuming noninteracting electrons, in the absence
of the intrinsic spin-orbit coupling, one can independently consider
the spin species with opposite polarization and then relate the spin
Hall effect to the anomalous Hall effect. \cite{ehr} In particular,
at magnetic field $B=0$, we find $d\SHtensor/d\EFermi=(\hbar/2\mu e)\, d\AHtensor/dB|_{B=0}$,
with $\AHtensor=\frac{1}{2}\left[\conductivityTensor(B)-\conductivityTensor(-B)\right]$,
and where $\conductivityTensor(B)$ is magnetic field-dependent charge
conductivity, $\EFermi$ is the Fermi energy, $\mu$ is the magnetic
moment, and $e$ is the electron charge. From the Onsager relation
for $\conductivityTensor(B)$, we see that $\SHtensor$ must be anti-symmetric.
Using this property for a homogeneous two- or three-dimensional system,
we find \begin{align}
\spingen & =-\frac{1}{2}\sum\nolimits _{ijk}\sigma_{ij}^{\mathrm{SH}}\epsilon_{ijk}(\nabla\times\vec{E})_{k}=0,\end{align}
i.e., the spin generation rate vanishes and no spin polarization would
be found far away from boundaries in the linear regime. 

For a system with intrinsic spin-orbit interaction, the interpretation
of the spin currents defined above becomes less clear. \cite{ehrReview,shi_spinCurrentDef}
Thus, instead we consider the electrically induced spin density $\vec{s}(\vec{r})$
inside the sample on length scales larger than $\Ls$. For typical
sample dimension $L>\Ls$, the field $\vec{E}$ changes slowly over
$\Ls$, and we make a gradient expansion linear in $E$, leading to
$s_{i}(\vec{r})\approx\sum_{j}A_{ij}E_{j}(\vec{r})\,+\sum_{jk}B_{ijk}\partial_{j}E_{k}(\vec{r}).$
For GaAs, the cubic symmetry group $\mathrm{T_{d}}$ contains mirror
planes, implying that $\vec{s}$ transforms as a pseudo-vector; thus
all $A_{ij}$ vanish. The only pseudo-vector linear in $\nabla$ and
$\vec{E}$ is $\nabla\times\vec{E}=0$. Going to the next order in
the gradient expansion, there are contributions $\propto(\partial_{xx}^{2}-\partial_{yy}^{2})E_{z}$
and $\propto\partial_{z}(\partial_{x}E_{x}-\partial_{y}E_{y})$, where
$x,\, y,\, z$ are the principal crystallographic axes. These terms
also vanish in the in-plane geometry ($\partial_{z}E_{i}=0$) with
an in-plane electric field ($E_{z}=0$).

We have found that spin generation away from boundaries may be forbidden
or strongly suppressed in the linear transport regime. Therefore,
we will now discuss how nonlinear effects can lead to such spin generation.
We will restrict ourselves to the case of extrinsic spin currents
that are generated by spin-dependent impurity scattering; they are
given by $\jSHz=\hat{\vec{z}}\times(\frac{\gamma}{2e}\jCharge-2\lambda n\frac{e}{\hbar}\vec{E})$,\cite{ehr}
where the two terms correspond to the skew-scattering and side-jump
contributions, respectively. Here, $\jCharge=\conductivityTensor\vec{E}$
is the charge current, but now we allow the conductivity tensor to
be a function of the electric field. Further, $\lambda$ is a material-dependent
spin-orbit coupling constant and the skewness $\gamma$ is of order
$\lambda$ but it also depends on the properties of scatterers and
on electron distribution function. The side-jump part of the current
does not contribute to $\spingen$, and we obtain \begin{equation}
\spingen=\frac{1}{2e}(\nabla\times\gamma\conductivityTensor\vec{E})_{z}.\label{eq:spinGen}\end{equation}
 Assuming that $\gamma$ is constant, that the conductivity is isotropic,
and that its position dependence is described by the local electric
field, $\conductivityTensor(\vec{r})=\sigma(E(\vec{r}))$, Eq.~(\ref{eq:spinGen})
simplifies to \begin{equation}
\spingen=\frac{\gamma}{2e}\:\frac{d\sigma}{dE}\,(\nabla E\times\vec{E})_{z}.\end{equation}
In samples where the extrinsic spin Hall effect was observed \cite{kato,sih}
the charge conductivity $\sigma(E)$ was found to be an increasing
function of field;\cite{kato,davidprivate} we expect that $\spingen$
is finite in these samples, and spin polarization can be generated
away from boundaries (see discussion below).

Next, we show how an $anisotropic$, nonlinear conductivity leads
to spin generation away from the boundaries. In general, according
to Eq.~(\ref{eq:spinGen}), inhomogeneous electric fields are required
for a finite $\spingen$. Thus, we now analyze the inhomogeneous field
in a Corbino geometry, where a total current $I$ is injected at $r=a$
into an infinite two-dimensional sample. In a (001) film of crystal
of full cubic symmetry, the leading nonlinearities in $\jCharge(\vec{E})$
are \begin{equation}
\jChargeComp{i}=(\sigma+\sigma_{2}E^{2})E_{i}+\sigma_{1}E_{i}^{3},\end{equation}
where the components $i=x,y$ are taken along the principal crystal
axes and we assume that $\sigma_{1,2}E^{2}(a)\ll\sigma$. One can
expect anisotropic nonlinear terms of considerable magnitude for many-valley
semiconductors like SiGe quantum wells, and also for Al$_{x}$Ga$_{1-x}$As
quantum wells with $x$ close to the direct-indirect gap transition.
We first solve for the electrostatic potential $\Phi(\vec{r})$; in
polar coordinates it will be of the form $\sum_{\mathrm{m}}\Phi_{\mathrm{m}}(r)\cos4\mathrm{m}\varphi$.
The term $\Phi_{0}(r)$ will not contribute to $\spingen$ {[}Eq.~(\ref{eq:spinGen})],
so we next consider the lowest harmonic, $\Phi_{1}(r)\cos4\varphi$.
Because $\divergence\jCharge=0$, we see that \begin{align}
 & \frac{d^{2}\Phi_{1}(r)}{dr^{2}}+\frac{1}{r}\frac{d\Phi_{1}(r)}{dr}-\frac{16}{r^{2}}\Phi_{1}(r)+\frac{3\sigma_{1}}{2\sigma r^{4}}\biggl(\frac{I}{2\pi\sigma}\biggr)^{3}=0.\label{eq6}\end{align}
 Requiring that $\Phi_{1}(r\to\infty)=0$, we obtain\begin{align}
 & \Phi_{1}(r)=\frac{\sigma_{1}}{8r^{2}}\biggl(\frac{I}{2\pi\sigma}\biggr)^{3}.\label{eq7}\end{align}
 The spin generation rate is then \begin{equation}
\Gamma_{z}(r,\varphi)=\frac{3\gamma}{4e}\sigma_{1}\biggl(\frac{I}{2\pi\sigma}\biggr)^{3}\frac{\sin4\varphi}{r^{4}}.\label{eq8}\end{equation}
 So indeed, the combined anisotropy and nonlinearity of conductivity
lead to a spin generation, which, for $r\gg L_{s}$, results in the
spin density is $s_{z}(r,\varphi)=\Gamma_{z}(r,\varphi)\,\taus\propto I^{3}\sin(4\varphi)\,/r^{4}$.
We emphasize that this polarization falls off only as a power law.
Furthermore, it consists of four sectors of up-spins separated by
four sectors of down-spins.

\begin{figure}[t]
\begin{center}\includegraphics[width=85mm,keepaspectratio]{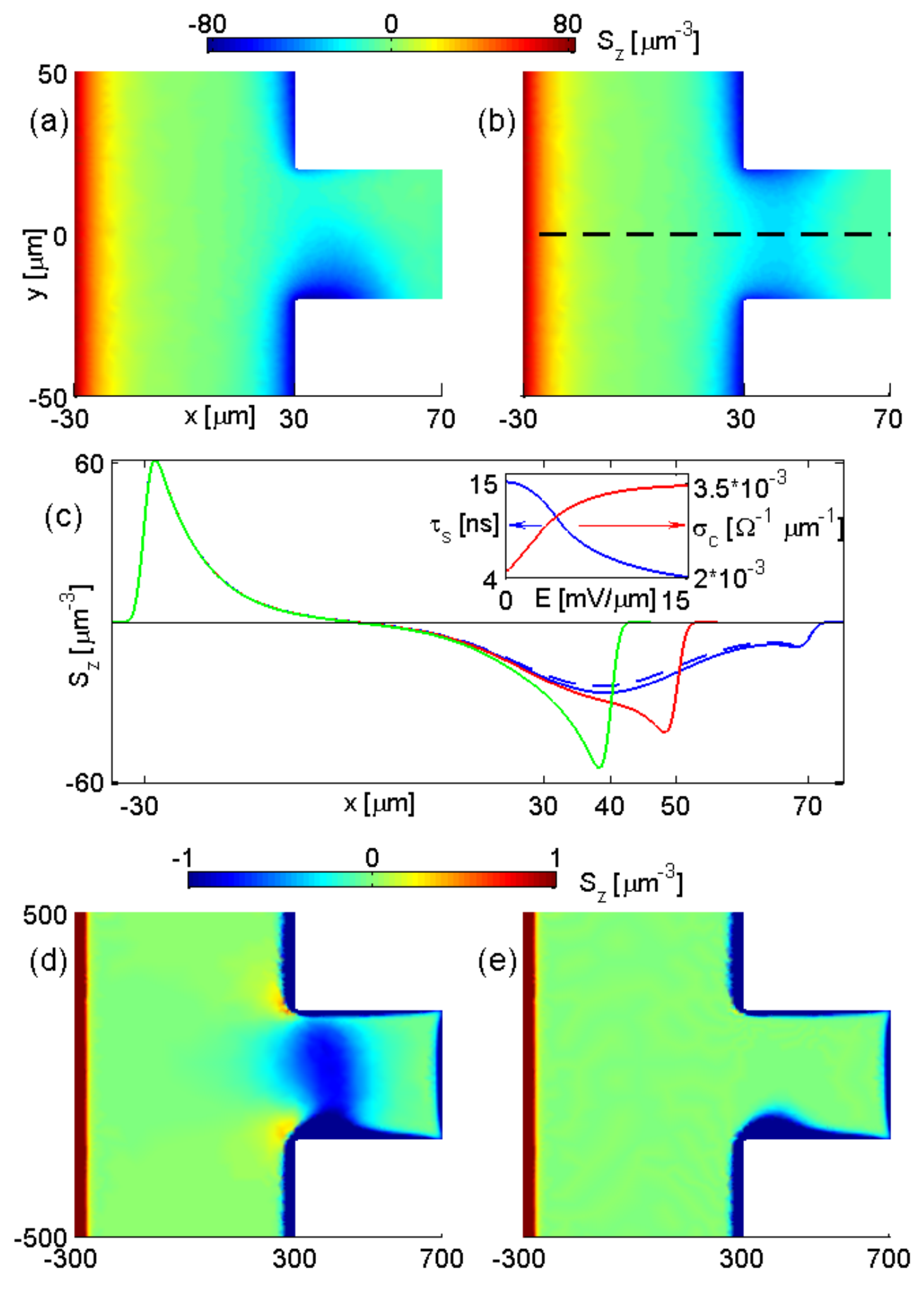}\end{center}

\caption{\label{fig:Tsolution} Simulated spin accumulation $s_{z}(\vec{r})$
for experimental geometry and parameters of Ref.~\onlinecite{sih}.
Electron flow is from bottom to top in the main channel, and spills
over into the side-arm. Distances are measured in microns and electric
field in the main channel is $9.5\:\mathrm{mV}/\mu\mathrm{m}$. (a)
Numerical solution of spin drift-diffusion equation {[}Eq.~(\ref{eq:kinetic})]
for a 40 $\mu\mathrm{m}$ sidearm in the dc regime. (b) Symmetrized
spin accumulation $\frac{1}{2}\left[s_{z}(V)-s_{z}(-V)\right]$. (c)
Spin polarization $s_{z}(x,y)$ (convoluted with a Gaussian with standard
deviation of 1$\mu\mathrm{m}$ that is associated with laser spot)
for $y=0$ (along dashed line in (b)) and for side-arms of depth 10
(green), 20 (red), and $40\:\mu\mathrm{m}$ (blue) is in good agreement
with the experimental data in Fig.~1c of Ref.~\onlinecite{sih}.
The dashed line shows a simulation without bulk spin generation ($\spingen=0$)
. The inset shows spin relaxation time $\taus(E)$ and charge conductivity
$\sigma(E)$ as used in our simulation. Panels (d), (e) show $s_{z}(r)$
on a different color scale for a sample larger by a factor $\alpha=10$.
In (e), the spin generation rate $\spingen$ is neglected.}
\end{figure}

We now analyze the spin polarization $s_{z}(\vec{r})$ in systems
with $isotropic$ (nonlinear) conductivity but with less symmetric
geometries. We consider experiments on unstrained GaAs samples by
Sih \emph{et al.}.\cite{sih} The spin Hall effects in such samples
is believed to be primarily extrinsic in origin. \cite{kato,ehr,tds}
In these experiments a T-shaped geometry, as shown in Fig.~\ref{fig:Tsolution},
was used. In an electric field, electrons flow from the bottom to
the top of the main channel, some of them entering into the side-arm;
the gradient of the electrical field $E(\vec{r})$ becomes large in
the region near the entrance of the arm, and a field-dependent conductivity
will then lead to spin generation across this region. Furthermore,
due to the spin Hall effect, spins are also generated near the sample
edges and can then diffuse and drift along the electric field into
the center of the side arm---below, we find that this latter mechanism
dominates the experimental observations. We choose realistic values
of the parameters as follows. Electron density $n=3\times10^{16}\:\mathrm{cm}^{-3}$
and sample dimensions are taken from Ref.~\onlinecite{sih}. Further,
we assume that the sample is homogeneous and the spatial dependence
of $\sigma$, $\taus$, and $\Ds$ is controlled by the local field
$E(\vec{r})$.  Low field conductivity $\sigma(E)$ for GaAs samples
was obtained from unpublished data,\cite{davidprivate} and for higher
fields we took the $E$-dependence of the conductivity measured for
$\mathrm{In}_{0.07}\mathrm{Ga}_{0.93}\mathrm{As}$ in Ref.~\onlinecite{kato}
as a guideline, which increased by a factor of two when the field
increased from 0 to 20 $\mathrm{mV}/\mathrm{\mu m}$.  The spin relaxation
time $\tau_{s}(E)$ was loosely based on the experimental data taken
at two points away from the boundaries for a range of electric fields;\cite{davidprivate}
we show our assumed $\sigma(E)$ and $\tau_{s}(E)$ in Fig.~1(c).
Furthermore, it was found in Ref.~\onlinecite{kato} that the spin
diffusion length $L_{s}$ was field-independent within error bars;
thus we take $D_{s}(E)=L_{s}^{2}/\tau_{s}(E)$ with constant $L_{s}$.
We chose $L_{s}=7\:\mu\mathrm{m}$ found from the best fit to the
data of Ref.~\onlinecite{sih} for the main channel. Finally, we
take spin-orbit coupling constant $\lambda=5.3\:\mbox{\AA}{}^{2}$
and estimate the skewness $\gamma=1/700$. \cite{ehr,skewness} In
our simulation, we first solve for the electrostatic potential $\Phi(\vec{r})$
for an applied dc voltage and determine the spin generation rate $\spingen(\vec{r})$
{[}Eq.~(\ref{eq:spinGen})] and the drift velocity $\vdrift(\vec{r})=\jCharge(\vec{r})/ne$.
Because there are no direct indications of a considerable spin relaxation
at the boundaries, we consider the spin-conservation boundary condition
$\hat{\vec{n}}\cdot(D_{s}\nabla s_{z}-\jSHz)=0$. This accounts for
the spin generated at the boundaries due to the spin Hall effect.
We then solve the spin drift-diffusion equation {[}Eq.~(\ref{eq:kinetic})]
and find the stationary spin polarization $s_{z}(\vec{r})$.

\begin{quote}
 
\end{quote}

The simulated $s_{z}(\vec{r})$ is shown in Fig.~\ref{fig:Tsolution}(a),
taking experimental values for sample dimensions and electric field.\cite{sih}
The most striking feature is a spin distribution inside the side-arm:
it results from spins generated at the boundaries near the corner
of the lower part of the side-arm, which diffuse and then drift a
distance $\Ldrift=\vdriftAbs\taus$, which is much longer than $L_{s}$.
Conversely, near the upper part of the side-arm the spin population
is very low, since spins generated at the boundary drift out of the
side-arm; i.e., the strong drift ($\Ldrift>\Ls$) leads to an asymmetric
spin distribution. However, in Ref.~\onlinecite{sih} a square wave
voltage $V$ was applied for lock-in detection; then the measured
spin polarization, $\frac{1}{2}\left[s_{z}(V)-s_{z}(-V)\right]$,
becomes symmetric, and we show the 'symmetrized' polarization in Fig.~\ref{fig:Tsolution}(b).
We find good agreement with experimental data, see Fig.~1(c) in Ref.~\onlinecite{sih}:
in particular, the maximum of spin distribution is at a similar position
inside the side-arm, about 10 - $15\:\mu\mathrm{m}$ deep. In Fig.~\ref{fig:Tsolution}(c),
we show $s_{z}$ along the central section of the side-arm for different
depths of the side-arm and find that for depths of 10 or 20 $\mu\mathrm{m}$,
the spin population is maximal at the right edge of the arm. 

The simulation includes contributions from spins generated away from
boundaries with rate $\spingen$ {[}Eq.~(\ref{eq:spinGen})]. For
comparison, we show  the spin polarization for $\spingen=0$, see
dashed line in Fig.~1(c). We, however, find that the absence of $\spingen$
changes the resulting spin polarization by less than 15\%. Therefore,
in the parameter regime of the experiments of Ref.~\onlinecite{sih},
the dominant contribution to the \textit{\emph{stationary}} spin distribution
inside the side-arm comes from the drift of the spins generated at
the sample boundaries. This results from the large drift distance
$\Ldrift$, which is of the same length scale as the width of the
side-arm. 

In order to distinguish spins generated in the bulk from those originating
at the boundaries, one needs a set up where both $\Ls$ and $\Ldrift$
are smaller than the width of the side arm, $L$. While $\Ls<L$ was
satisfied in the experiments of Ref.~\onlinecite{sih} and in the
simulations above, $\Ldrift<L$ was not. The inequality can be achieved
(i) by reducing $\Ldrift$ with a smaller applied field (and, therefore,
smaller drift velocity), (ii) by reducing $\Ldrift$ by pulsing the
applied fields and thus allowing drift only for a shorter time, (iii)
by using an alternating sequence of pulsed electric fields, which
can largely cancel the drift and the spin generation at the boundaries,
or (iv) by increasing sample dimensions $L$. This exponentially suppresses
the contributions from the boundaries (as long as $\Ls$ is sufficiently
short), making the relative contribution of the generation away from
boundaries arbitrarily large. However, this also suppresses the absolute
value of spin polarization, as we discuss now.

For approach (i), a smaller drift velocity is achieved by choosing
a sufficiently small electric field, so that $\mathrm{\vdriftAbs}\ll L/\taus$.
Because of the scaling $\spingen\propto\sigma'(E)E^{2}/L$, with $\sigma'(E)=d\sigma/dE$,
small drift velocities also imply that spin polarization due to generation
away from boundaries is weak. (ii) When the electric field is only
applied for pulses of duration $\tPulse$, which are short compared
to $\taus$, the drift length is reduced to $\Ldrift=\tPulse\vdriftAbs$.
Also, the pulses must be at least $\taus$ apart, thus the average
spin polarization signal is reduced from the dc case by a factor smaller
than $\tPulse/\taus$. For example, applying electric field pulses
with $\tPulse=1\:\mathrm{ns}$ and separation time $20\:\mathrm{ns}$,
for the parameters used in Fig.~\ref{fig:Tsolution}, spin generation
away from boundaries could be observed with a time-averaged value
of $s_{z}\lesssim0.2\,\mu\mathrm{m}^{-3}$. Another approach (iii)
is to use alternating pulses of large positive $E$ for duration $t_{1}$
and negative $E$ of smaller magnitude for a longer time $t_{2}$,
such that the time-averaged electric current is zero. Then, if the
period $t_{1}+t_{2}$ is smaller than both $\tau_{s}$ and $L/\mathrm{v_{dr}}$,
the effects of spin drift should be largely canceled. Furthermore,
during $t_{1}$ and $t_{2}$ equal amounts of spin polarization are
generated at the boundary, but with opposite signs, which leads to
a cancellation of the contribution of spins generated at the boundaries.
In contrast, because the bulk spin generation $\spingen$ is nonlinear,
it can have a nonzero time average. Thus, the effectiveness of this
scheme will depend on the strength of nonlinearities involved.

Note that in approaches (i)-(iii), diffusion of spins generated at
the boundaries may still dominate $s_{z}(\vec{r}$) in the bulk. For
example, in simulations with the geometry of Fig.~\ref{fig:Tsolution}
but for smaller fields, the bulk $s_{z}$ still contains a large relative
contribution from spins generated at the boundaries. To eliminate
such a diffusion effect, one could use approach (ii) in a pump-probe
scheme, where the spin polarization is detected shortly after the
pulse, providing a direct measurement of $\spingen(\vec{r})$.

The most straightforward way to reveal spin generation away from boundaries
is to increase the sample's linear dimensions by a factor $\alpha$
big enough so that the sidearm opening exceeds the drift length $L_{dr}$,
while keeping the average electric field in the main channel fixed
(iv). This reduces $\Gamma_{z}$ and hence the bulk spin polarization
by a factor $\alpha$.  However, the signal to noise   should
not decrease since the signal can be taken from an area that is larger
by a factor of $\alpha^{2}$, thus reducing the noise by $\alpha$.
 Figs.~1(d,e) demonstrate spin populations in a sample 10 times
bigger than the sample of Fig.~\ref{fig:Tsolution}. A noteworthy
feature is found in the vicinity of the corners: the spin population
there is of the opposite sign than the population on the adjacent
boundary. This is a distinctive feature of spin generation away from
boundaries. For comparison, we show in Fig.~1(e) the spin polarization
for $\spingen=0$.  No spin polarization is found in the central
part of the side-arm; therefore, the spin polarization in the central
part  of Fig.~1(d) indeed results from spins generated away from
boundaries.

In conclusion, we have shown that, in a linear transport regime, the
generation of spin polarization away from boundaries is forbidden
or strongly suppressed. However, in nonlinear transport, spins can
be generated away from boundaries and we analyze such generation resulting
from spin Hall currents due to the {}``extrinsic'' skew scattering
mechanism. For an anisotropic nonlinear charge conductivity tensor,
we analytically evaluate the spin generation in a Corbino geometry.
We also simulate the spin polarization in a T-shaped geometry using
isotropic nonlinear conductivity and field-dependent spin lifetimes,
appropriate to recent experiments. We find that the spin accumulations
in the experiments of Ref.~\onlinecite{sih} were primarily due to
drift of spins generated at the boundaries, but we suggest other setups,
where drift should be relatively unimportant, and nonlinear spin generation
away from boundaries should be observable.

We thank D.D. Awschalom for helpful discussions and for providing
us with unpublished data. This work was supported in part by NSF grants
DMR05-41988 and PHY01-17795, and by Harvard Center for Nanonscale
Systems.

Note Added: In a recent work, Stern $et\, al.$ \cite{davidprivate,stern}
have independently shown that drift alone can account for observations
of Ref. \onlinecite{sih}. Golizadeh-Mojarad and Datta \cite{datta}
have obtained spin polarizations qualitatively similar to those seen
in experiments \cite{sih} using a model with intrinsic (Rashba) spin
orbit coupling. However, the spin coherence length and sample dimensions
used were two orders of magnitude smaller than the experimental values.
\textbf{}Pershin and Di Ventra \cite{pershin} considered rectangular
samples with inhomogeneous charge density $n$, which leads to spin
generation  linear in the electric field, assuming that $\sigma\propto n$
{[}cf.~Eq.~(\ref{eq:spinGen})].

\clearpage

\begin{thebibliography}{10}
\bibitem{dp71}M.I. Dyakonov and V.I. Perel, Phys. Lett. \textbf{35A},
459 (1971). 

\bibitem{hirsch}J.E. Hirsch, Phys. Rev. Lett. \textbf{83}, 1834 (1999). 

\bibitem{kato}Y.K. Kato, R.C. Myers, A.C. Gossard, and D.D. Awschalom,
Science \textbf{306}, 1910 (2004).

\bibitem{murakami}S. Murakami, N. Nagaosa, and S.-C. Zhang, Science,
\textbf{301}, 1348 (2003).

\bibitem{sinova}J. Sinova, D. Culcer, Q. Niu, N.A. Sinitsyn, T. Jungwirth,
and A.H. MacDonald, Phys. Rev. Lett., \textbf{92}, 126603 (2004).

\bibitem{wunder}J. Wunderlich, B. Kaestner, J. Sinova, and T. Jungwirth,
Phys. Rev. Lett. \textbf{94}, 047204 (2005). 

\bibitem{ehr}H.-A. Engel, B.I. Halperin, and E.I. Rashba, Phys. Rev.
Lett. \textbf{95}, 166605 (2005). 

\bibitem{tds}W.-K. Tse and S. Das Sarma, Phys. Rev. Lett. \textbf{96},
056601 (2006). 

\bibitem{ehrReview}For theory reviews, see H.-A. Engel, E.I. Rashba,
and B.I. Halperin, cond-mat/0603306; J. Schliemann, Int. J. Mod. Phys.
B \textbf{20}, 1015 (2006) and references therein. 

\bibitem{sih}V. Sih, W.H. Lau, R.C. Myers, V.R. Horowitz, A.C. Gossard,
and D.D. Awschalom, Phys. Rev. Lett. \textbf{97}, 096605 (2006). 

\bibitem{davidprivate}D.D. Awschalom (private communication).

\bibitem{skewness} This value of $\gamma$ is found from a corrected
numerical evaluation, following Ref.~\onlinecite{ehr}. 

\bibitem{shi_spinCurrentDef}J. Shi, P. Zhang, D. Xiao, and Q. Niu,
Phys. Rev. Lett. \textbf{96}, 076604 (2006). 

\bibitem{stern}N.P. Stern \emph{et al.} (in preparation, to be submitted
to Appl. Phys. Lett.)

\bibitem{datta}R. Golizadeh-Mojarad and S. Datta, cond-mat/0703280.

\bibitem{pershin}Y. Pershin and M. Di Ventra, cond-mat/0703310.

\end{thebibliography}
\end{document}